\documentstyle[11pt,newpasp,epsf,twoside]{article}
\def\etal{\mbox{\it et al.}}
\def\kms    {\ifmmode{{\rm ~km~s}^{-1}}\else{~km~s$^{-1}$}\fi}
\font\cmss=cmss10 scaled 1200

\def\edcomment#1{\iffalse\marginpar{\raggedright\sl#1\/}\else\relax\fi}
\reversemarginpar

\begin{document}
\title{Accretion and Outflow Traced by H$_2$O Masers in the Circinus AGN}
 \author{L. J. Greenhill, J. M. Moran}
\affil{Harvard-Smithsonian Center for Astrophysics, 
       60 Garden St, Cambridge, MA 02138 USA}

\author{R. S. Booth}
\affil{Onsala Space Observatory, SE-43992 Onsala, Sweden}

\author{S. P. Ellingsen, P. M. McCulloch}
\affil{University of Tasmania, Discipline of Physics, 
       GPO Box 252-21, Hobart, TAS 7001 Australia}

\author{D. L. Jauncey, R. P. Norris, J. E. Reynolds, A. K. Tzioumis}
\affil{Australia Telescope National Facility, 
       PO Box 76, Epping, NSW 2121 Australia}

\author{J. R. Herrnstein}
\affil{Renaissance Technologies, 
       600 Route 25A, E. Setauket, NY 11733 USA}

\begin{abstract}
The first VLBI images of H$_2$O maser emission in the Circinus Galaxy AGN show
both a warped, edge-on accretion disk and an outflow 0.1 to 1 pc from the
central engine.  The inferred central mass is $1.3\times10^6$ M$_\odot$, while
the disk mass may be on the order of $10^5$ M$_\odot$, based on a nearly 
Keplerian rotation curve. The bipolar, wide-angle outflow appears to contain
``bullets'' ejected from within $<0.1$ pc of the central mass. The positions of
filaments and bullets observed in the AGN ionization cone on kpc-scales
suggest that the disk channels the flow to a radius of $\sim 0.4$ pc, at which
the flow appears to disrupt the disk.
\end{abstract}

\section{Introduction}

At a distance of $\sim 4$~Mpc (Freeman\ \etal\ 1977) the Circinus galaxy is
one of the nearest Seyfert\,II galaxies that hosts an extragalactic H$_2$O
maser. The $3\times10^{42}$ erg s$^{-1}$ (2-10 keV) central engine is obscured
at energies below 10 keV by a large gas column, n$_{\rm H}\sim4\times10^{24}$
cm$^{-2}$ (Matt\ \etal\ 1999). Veilleux \& Bland-Hawthorn (1997) observed
linear optical filaments and compact knots in the ionization cone, which opens
by at least $90^\circ$  at a mean position angle (PA) of roughly $290^\circ$.
The AGN also drives a kpc-scale nuclear outflow that creates 
bipolar radio lobes at PA $\sim 295^\circ$ (Elmouttie\ \etal\ 1998),
largely along the rotation axis of a nuclear $^{12}$CO ring (Curran\
\etal\ 1998).

There is good evidence that extragalactic H$_2$O masers in NGC\,4258 and
NGC\,1068 trace edge-on accretion disks bound by central engines $\ga 10^6$
M$_\odot$.  Discovery of maser emission in Circinus more or less symmetrically
bracketting the systemic velocity led to speculation that these masers 
also trace an accretion disk (Greenhill\ \etal\ 1997). Though the detection
of an accretion disk reported here confirms that prediction, the additional
discovery of qualitatively new, nondisk emission introduces a fresh element
in modeling the AGN dynamics.

\section{Observations and Data}

We observed the Circinus H$_2$O maser three times in 1997
(June \& July) and 1998 (June) with four stations of the ATNF 
Long Baseline Array. We used a spectral-channel separation of 0.21\kms~
over 437\kms~in 1997 and 553\kms~in 1998, centered on the known emission.
We calibrated the data with standard VLBI techniques and estimated a new
astrometric position for the 565\kms~line (where we adopt the radio 
astronomical definition of Doppler shift) by analyzing the time variation in
fringe rate. 
The new position, $\alpha_{2000}=14^h13^m09\rlap{.}^s95\pm0\rlap{}^s.02$,
$\delta_{2000}=-65^\circ20'21\rlap{.}''2\pm0\rlap{.}''1$ is the best estimate
so far of the AGN.

We calibrated tropospheric path-length fluctuations by self-calibrating the
$\sim 565\kms$~emission and achieved 0.025 - 0.045 Jy noise ($1\sigma$) in the
deconvolved synthesis images, depending on the epoch. We fitted a 2-D Gaussian
model brightness distribution to each spot stronger than $6\sigma$. Because
spectral lines come and go from epoch to epoch, we superposed the maps for all
epochs, so as to trace the underlying dense molecular gas as completely as
possible with the available data (Figure\,1).

\section{The Warped Disk and Outflow}

At each epoch, the sky distribution of maser emission comprises three
populations: (1) a thin, gently curved, {\cmss S}-shaped locus of highly
redshifted and blueshifted emission arcs to the southwest and northeast,
respectively (aka ``high-velocity'' emission),  (2) emission close to the
nominal systemic velocity of the galaxy that lies between the high-velocity
arcs (aka ``low-velocity'' emission), and (3) modestly Doppler shifted
emission broadly distributed in knots that lie north and west (redshifted) and
south and east (blueshifted) of the low-velocity emission. In 1998 detectable
emission lay between 215\kms~and 677\kms.

We have fitted a model edge-on disk with smoothly varying position angle as a
function of radius (Figure\,1).  The observed disk inner and outer radii are
$\sim 0\rlap{.}''005$ and $\sim 0\rlap{.}''02$, respectively (0.1 - 0.4 pc).
The peak orbital speed is 237\kms, and the mass enclosed with 0.1 pc is
$1.3\pm0.1\times10^6$ M$_\odot$, for circular motion and a 451\kms~model
systemic velocity. Among the redshifted high-velocity masers, the peak
rotation velocity as a function of impact parameter from the dynamical center,
$b$, traces a rotation curve that declines as approximately $b^{-0.40\pm0.02}$
(Figure\,2).  We infer a disk mass on the order of $10^5$ M$_\odot$ between
0.1 and 0.4 pc.  The inner disk is edge-on, but if it is inclined toward the
outer radius, then the disk mass is larger.  An estimate of the Toomre
Q-parameter ($\la 0.4$) suggests that self-gravity in the disk is important, 
which is consistent with apparent clumpy substructure in the distribution of
disk emission.

The disk model derives its strongest support from the following observations:
(1) the angular distribution of high-velocity masers is highly elongated, roughly
symmetric, and perpendicular to the approximate axis of the known ionization
cone, (2) the apparent rotation curve is nearly Keplerian, (3) the innermost
red and blue-shifted emission and the low-velocity emission are colinear in
position and position-velocity space. The latter item indicates the
low-velocity emission lies close to the inner radius of the disk. 

Outflow is traced by emission between 300 and 450\kms, south and east of
the dynamical center of the disk, and by emission between 450 and 580\kms,
north and west of the dynamical center. The distribution of masers on the sky
suggests a wide-angle flow, and the segregation of red and blueshifted
emission suggests an inclined flow. Because H$_2$O maser emission requires
$n_{{\rm H}_2}\ga 10^9$ cm$^{-3}$, the masers probably represent high-density
``bullets'' immersed in a thinner, photoionized medium, equivalent to a
narrow-line region. However, the opacity of that region probably obscures all
but the nearside of the flow, and if the disk is inclined toward its outer
radius, then ionization along the disk surface probably obscures one side of
the flow preferentially.

The outflow appears to arise in the vicinity of the central
engine ($<0.1$ pc) rather than from the ``maser-disk'' by magneto-centrifugal
processes (Emmering\ \etal\ 1992; Kartje, K\"onigl, \& Elitzur 1999).
Notably, rotation is absent from the line-of-sight velocity and position data,
even for masers quite close to the disk (i.e., height/radius $\ll1$). Though
the outflow could originate in a ``hot'' disk that lies within the (warm)
molecular maser-disk, the physical conditions that support maser
emission are consistent with adiabatic expansion of AGN broad line clouds,
i.e., $10^{10}$ - 10$^{12}$ cm$^{-3}$ and of $10^4$ K (Brotherton\ \etal\
1994).

The disk structure disappears at a radius of $\sim 0.4$ pc (Figures\,1 \& 2),
perhaps because the wind disrupts the accretion disk, whose mean surface
density presumably declines with increasing radius. The existence of an outer
radius is supported circumstantially by two observations.  First,
outflow-borne masers all lie at the edge of or outside the shadow cast by the
truncated disk.  (We assume that the maser excitation depends on
irradiation and heating of molecular gas by hard X-rays, as in Neufeld,
Maloney, \& Conger (1994)). Second, the position angles at which the outflow
is free of occultation by the truncated disk coincide with the limits of the
kiloparsec scale ionization cone observed west of the nucleus (Veilleux \&
Bland-Hawthorne 1997). For example, the southern edges of the redshifted maser
outflow and the ionization cone, both lie at position angles of $\sim
-120^\circ$.  In addition, we note that the mean axis of the maser outflow,
$-52^\circ$, corresponds well to the orientation of the dominant [O\,III]
filament, $\sim -50^\circ$ and a radio hotspot (Elmouttie\ \etal\ 1998).

\begin{figure}[ht]
\plottwo{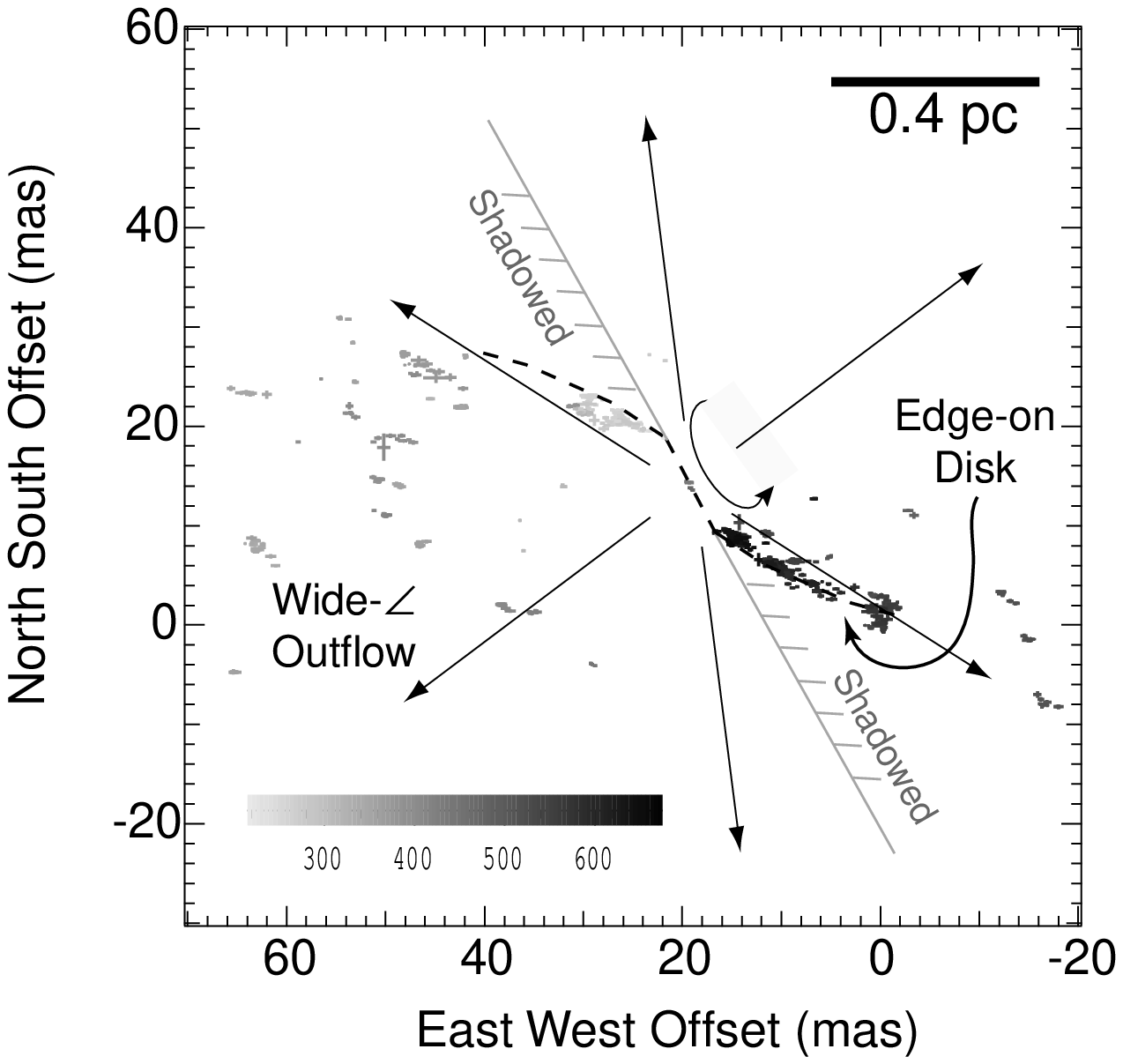}{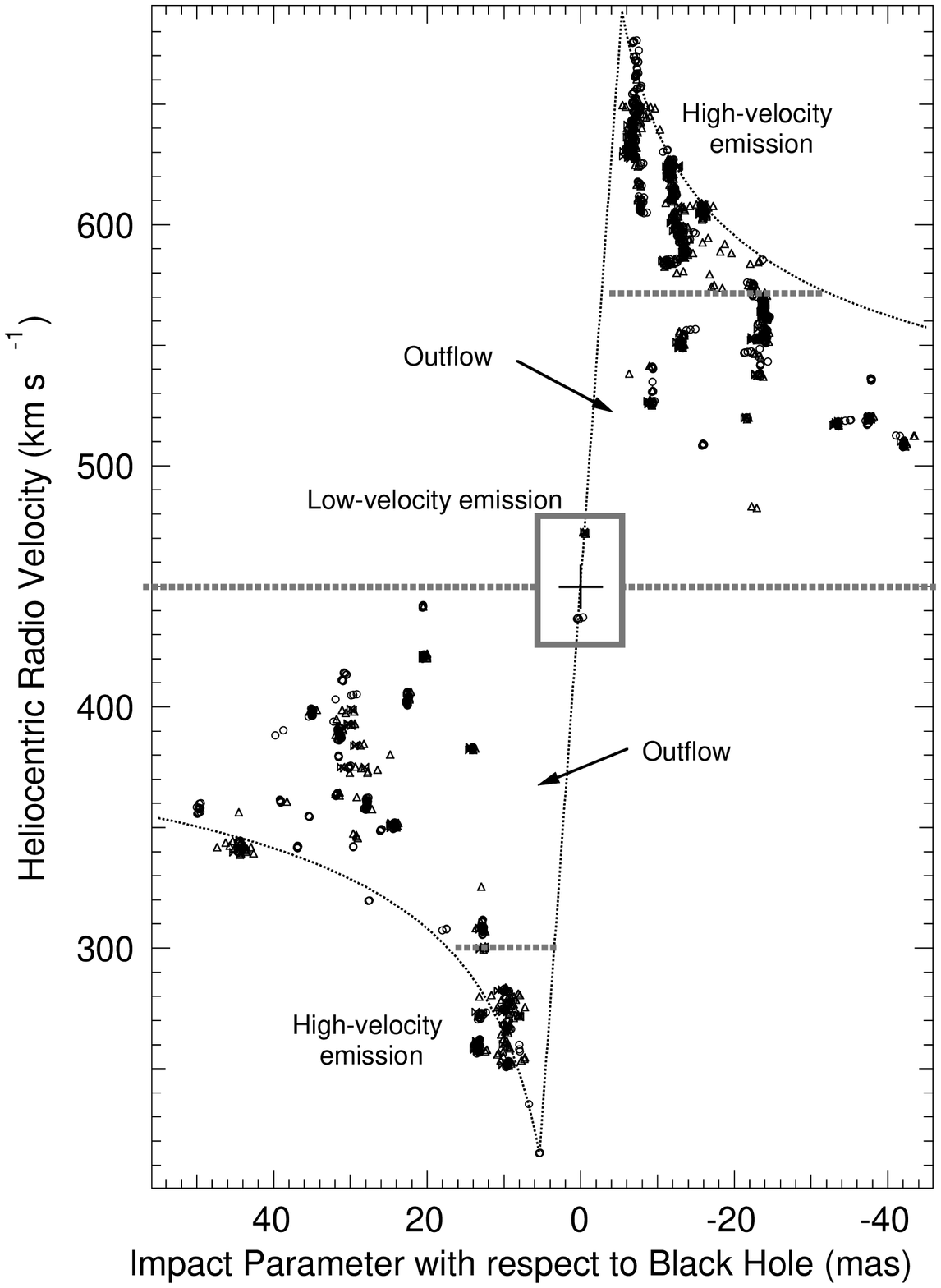}
\caption{ {\sl(left)} A model of the  proposed disk-outflow structure in the
inner 1 pc of the AGN overlayed on the combined distributions of maser spots
observed at three epochs (registered with 0.2 mas accuracy). Shading of spots
indicates heliocentric radio velocity, and error bars indicate total ($1\sigma$)
position uncertainties. Dashed lines trace the warped edge-on disk.  Outward
facing arrows indicate the wide-angle outflow.   The disk shadows some
regions, which cannot support maser emission probably because 
they are not irradiated by the central engine.  
Synthesized beams appear in the lower left corner.
{\sl (right)} Position-velocity diagram annotated to highlight the 
proposed disk and
outflow components.  Dashed rotation curves correspond to the midline of the
disk.  The steep diagonal line corresponds to the near side of the disk, at the
inner radius.  The outflow is Doppler shifted with respect to the
systemic velocity by on the order of $\pm100\kms$. }

\end{figure}

\end{document}